\documentclass[9pt,conference]{IEEEtran}
\usepackage{amssymb,amsthm,amsmath,array}
\usepackage{graphicx}
\usepackage[caption=false,font=footnotesize]{subfig}
\usepackage{xspace}
\usepackage[sort&compress, numbers]{natbib}
\usepackage{stmaryrd}
\usepackage{xcolor}
\usepackage{mathtools}
\usepackage{float}
\usepackage{textcomp}
\usepackage{algorithm}
\usepackage{lipsum}
\usepackage[spanish,german,english]{babel}
\usepackage{bbding}
\usepackage[T1]{fontenc}

\begin{document}
\title{A Blender-based channel simulator for FMCW Radar\\
\thanks{This work was supported by the Luxembourg National Research Fund (FNR) through the BRIDGES project MASTERS under grant BRIDGES2020/IS/15407066.}
}
\author{Yuan Liu,~\IEEEmembership{Student~Member,~IEEE,}
        Moein AHMADI,~\IEEEmembership{Member,~IEEE,}
        Johann Fuchs,
        Mohammad Alaee-Kerahroodi,~\IEEEmembership{Member,~IEEE,}
        M. R. Bhavani Shankar,~\IEEEmembership{Senior~Member,~IEEE} \\
\IEEEauthorblockA{Interdisciplinary Centre for Security, Reliability and Trust (SnT), University of Luxembourg, L-1855, Luxembourg\\
 Email: \{yuan.liu, moein.ahmadi, johann.fuchs, mohammad.alaee, bhavani.shankar \}@uni.lu}
}
\maketitle
\begin{abstract}
Radar simulation is a promising way to provide data-cube with effectiveness and accuracy for AI-based approaches to radar applications. 
This paper develops a channel simulator to generate frequency-modulated continuous-wave (FMCW) waveform multiple inputs multiple outputs (MIMO) radar signals. 
In the proposed simulation framework, an open-source animation tool called Blender is utilized to model the scenarios and render animations. 
The ray tracing (RT) engine embedded can trace the radar propagation paths, i.e., the distance and signal strength of each path.
The beat signal models of time division multiplexing (TDM)-MIMO are adapted to RT outputs. 
Finally, the environment-based models are simulated to show the validation. 
\end{abstract}

\section{Introduction}
\label{sec:intro}
Radio-based sensing systems have long and widely been used for various radar applications.
A basis for designing and optimizing sensor systems is the knowledge of radar channel characteristics \cite{yin2016propagation}. 
Conventionally channel models can be obtained by field measurements. However, measurement can be time-consuming and also expensive\cite{9321507}. The deterministic ray tracing (RT) \cite{6942178,8438326} has been used in wireless communication simulations. 
Following the trends, recent studies utilize the RT tool embedded in the animation software, e.g., Blender and Optix, for radar channel simulation \cite{8008254}. 
The Blender and the OptiX tools can model the dynamic scenarios and then render each frame of the animation, therefore capturing any slight motions of the target. 

However, there exist gaps in the state of the art. 
On the one hand, many rendering-based radar simulation works are for outdoors because they are motivated by the automatic driving industry, e.g., the OptiX-based ones \cite{9533181} and commercial simulator FEKO \cite{Feko}. On the other hand, most of them are short of considering dynamic simulation of the targets \cite{8008254,9533181,dageneral}.  
In this paper, a Blender-based mmWave MIMO radar simulator is developed for indoor applications built on the free and open-source Blender animation software. 
%
\section{Simulation Outline and Signal model} \label{Sec:simu_chain}
\begin{figure}
    \centering
\includegraphics[width=0.5\textwidth]{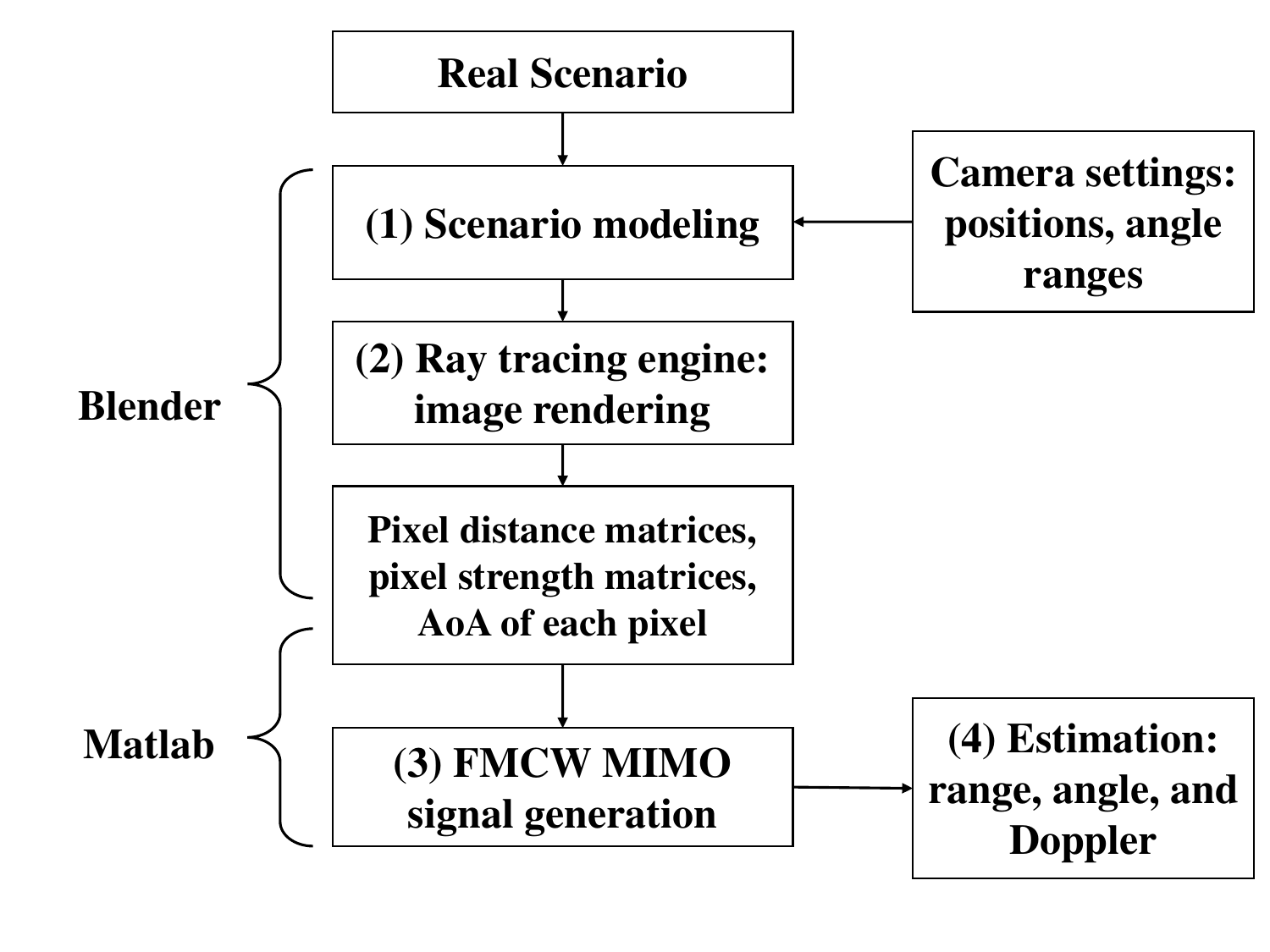}
    \caption{Outline of the simulation procedure. 
    }\label{fig:simu_1}
\end{figure}
The overall procedure of producing the FMCW radar signals using the RT tool embedded in Blender is illustrated in Fig.\ref{fig:simu_1}, which mainly consists of four steps, including scenario modeling, image rendering, signal generation, and target estimation. 

In scenario modeling, the tool is flexible to define the material properties, geometry size, and moving trajectories of the objects in the scenario by various means, e.g., the animation tools in Blender, add-on custom, and Python scripts via open interfaces.
The positions of the light source, camera, and camera angle ranges are in line with the positions of the transmitter (Tx), receiver (Rx) antennas, and antenna patterns, respectively. 
By the rendering engine chosen, e.g., the Cycles engine, each frame of the dynamic animation is rendered into a 2D picture, with each picture represented by a certain number of pixels, where the distance, strength, and angle of arrival (AoA) of each pixel can be obtained either directly by Blender or calculation. The frame rate and the number of pixels are defined manually. 
Using the Blender outputs, we can generate FMCW radar signals. 
 
Considering a $M$-Tx and $N$-Rx radar array. 
The beat signal of time division multiplexing (TDM)-MIMO in practice is \cite{9928573}
\begin{equation} 
\begin{aligned} \label{eq:tdm_10}
    \mathbf{h}_{T,n}&(n_s;l)  = \\
    &\sum_{m=1}^{M}  \sigma_{m,n,l}\exp (j2\pi(   \dfrac{2 \mu R_{m,n}}{c}\dfrac{n_s-1}{F_s} -f_{D}(l-1)T_b  )),
\end{aligned}
\end{equation}
where $m = 1, 2,..., M$ is the indices of Tx antennas, $n = 1, 2,..., N$ is the indices of Rx antennas, $l = 1, 2,..., L$ is the indices of TDM block, where one TDM block contains $M$ orthogonal FMCW chirps, $\sigma_{m,n,l}$ contains the attenuation of the RCS of the target and the propagation channel between the $m$th Tx and the $n$th Rx at the $l$th TDM block, $\mu$ is the slope of FMCW chirp, $R_{m,n}$ is the range, $c$ is the propagating velocity of the light, $n_s = 1, 2,..., N_s$ is the indices of fast time, $F_s$ is the sampling frequency, $f_D$ is the Doppler frequency shift, and $T_b$ is the chirp duration.

In Blender, a 3D scene is rendered into a 2D image, represented by $N_\text{az} \times N_\text{el}$ pixels. Each pixel can be regarded as one target, with the range and strength information being the rendering outputs of Blender. 
Based on (\ref{eq:tdm_10}) the beat signal of the $n$th Rx used in the simulation can be represented by $\mathbf{H}_{T_n} \in \mathbb{C}^{L \times N_s }$, where $L$ and $N_s$ denote the slow time and fast time axis, respectively, and the $(l,n_s)$-th entry can be expressed as
\begin{equation}
\begin{aligned} \label{eq:tdm_10_1}
    {{H}_{T_n}}_{l,n_s} &= \sum_{n_\text{az}=1}^{N_\text{az}}\sum_{n_\text{el}=1}^{N_{el}} \sum_{m=1}^{M}   P_{r_{n_{\text{frame}},n_{\text{el}},n_{\text{az}}}} \exp ( j2\pi(  \\
    & \dfrac{2 \mu R_{n_\text{frame}, n_\text{el},n_\text{az}} } {c}\dfrac{n_s-1}{F_s}   -2 \dfrac{f_l  V_{n_\text{fame},n_\text{el},n_{az} }}{c} \dfrac{l-1}{L} T_b  \\& + \dfrac{ ((m-1)N +n-1) \Delta d }{\lambda} \sin{ \theta_{n_\text{el},n_\text{az}}} ) ) , \\
\end{aligned}
\end{equation}
where $n_\text{az} = 1, 2,..., N_\text{az}$ is the indices of pixels at the azimuth axis, $n_\text{el} = 1, 2,..., N_\text{el}$ is the indices of pixels at the elevation axis, $n_\text{frame}$ is the indices of Blender frames, $R_{n_\text{frame}}$ is the rendering results of range matrix, Blender outputs $R_{n_\text{frame}, n_\text{el},n_\text{az}}$, $P_{{r_{n_\text{el},n_\text{az}} }}$, $\Theta_{n_\text{el},n_\text{az}}$, and $V_{n_\text{fame},n_\text{el},n_{az} }$ denote the range, strength, angle of arrivals and velocity, respectively, $\Delta d = \lambda/2$ is the interval between Rx antennas and $\lambda$ is the wavelength.

%

\section{Simulation and Discussion}\label{Sec:validation}
\begin{figure}[t]
    \centering
    \subfloat[Model of a human walks in a circle track.]{\includegraphics[width=0.4\textwidth]{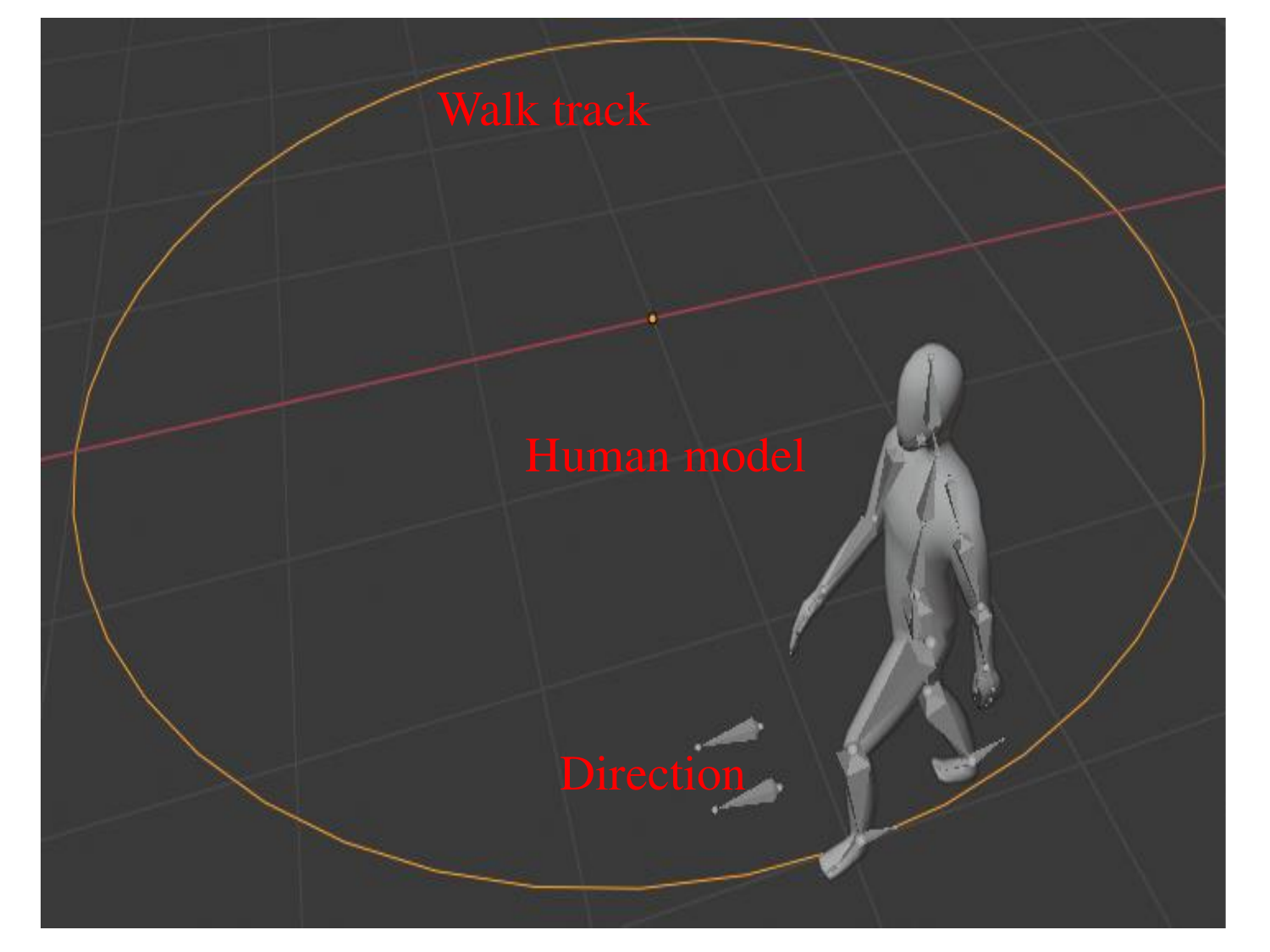}}
    \hfill
    \subfloat[Illustration of location of the sensor and pedestrian track.] {\includegraphics[width=0.47\textwidth]{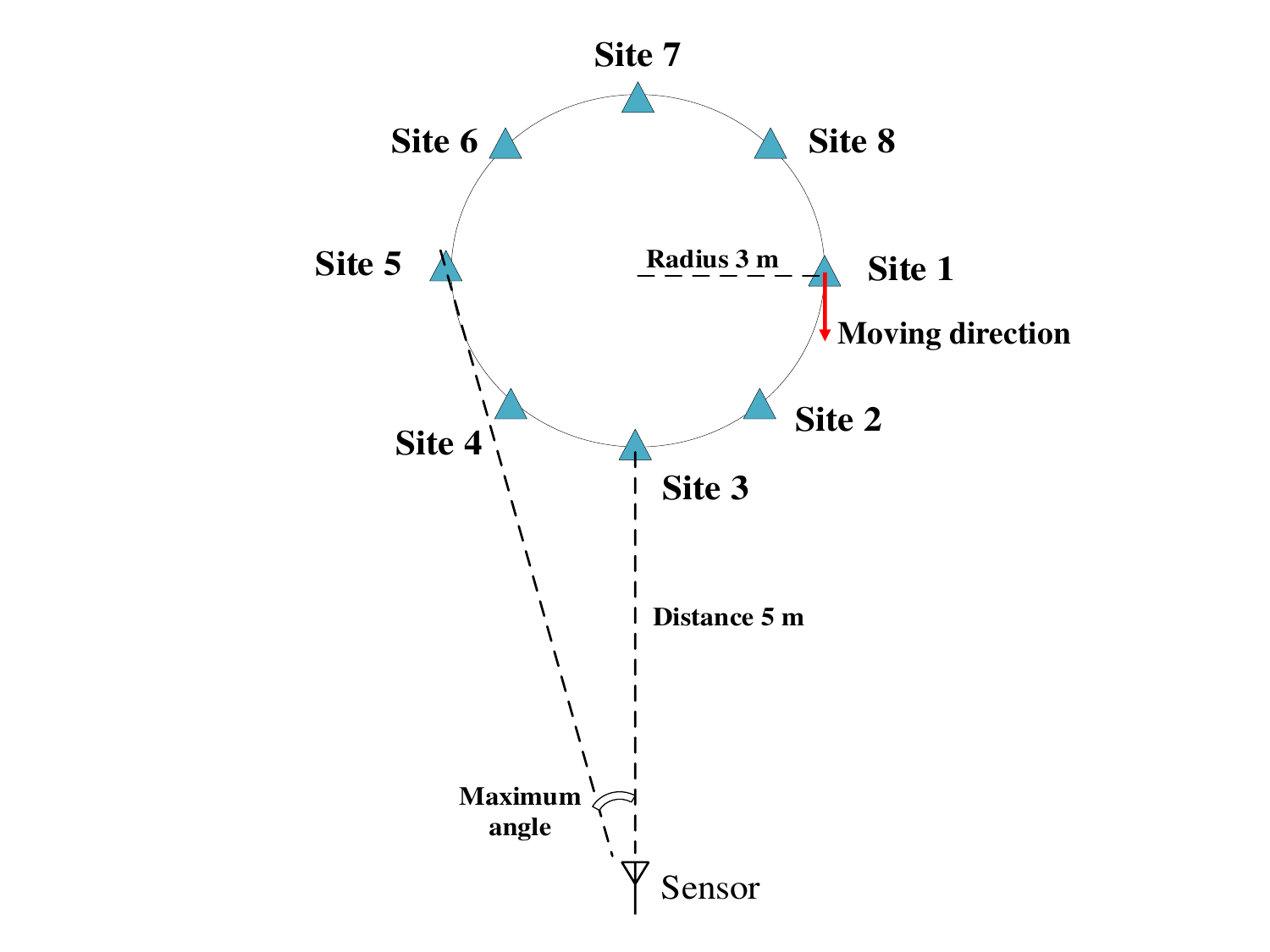}}
    \hfill
        \caption{Scenario.}
        \label{fig:scenario}
\end{figure}
\begin{figure}[t]
    \centering
\includegraphics[width=0.47\textwidth]{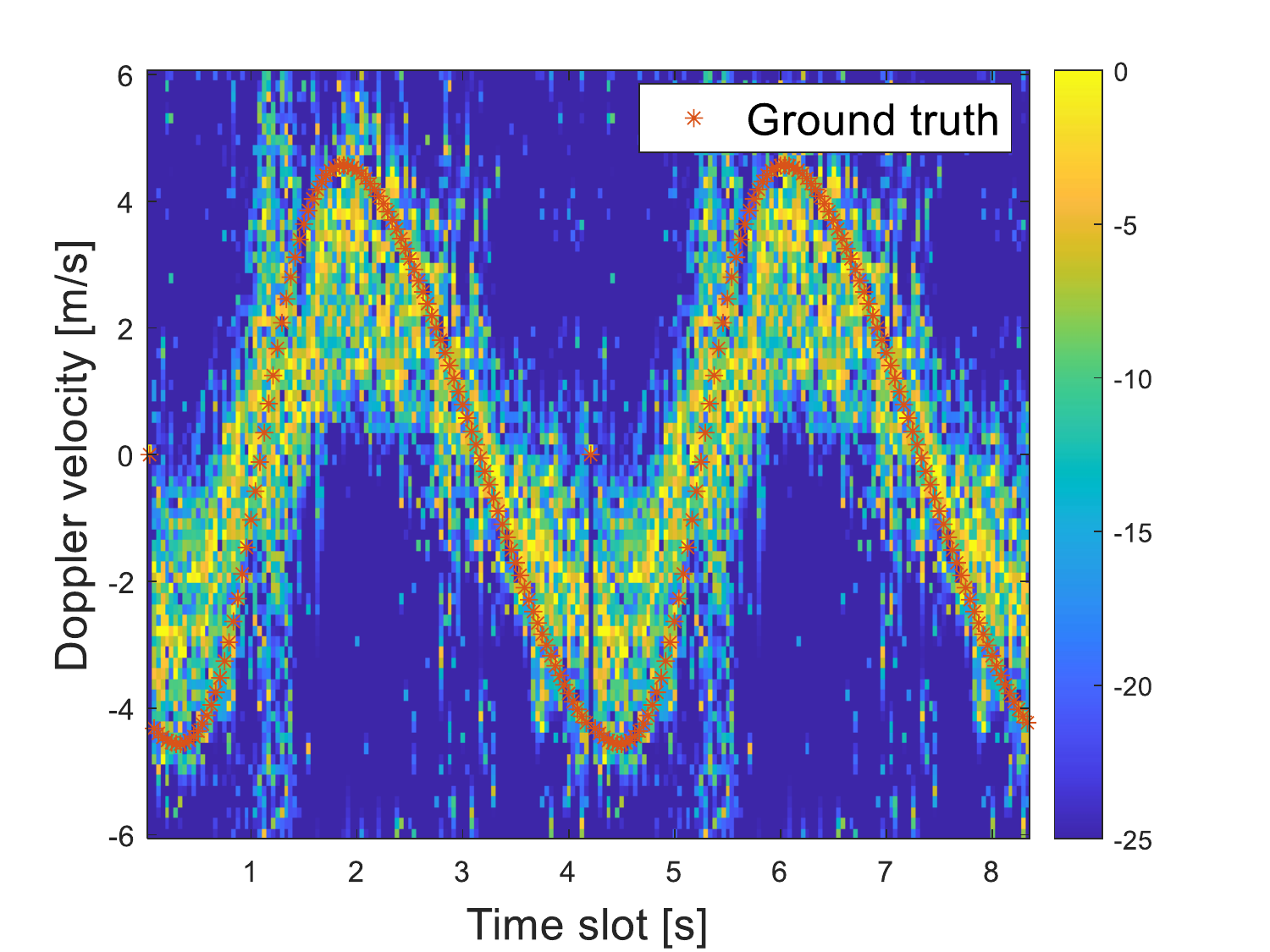}
    \caption{Doppler time plot: the envelope of the Doppler velocity changes like a sinusoidal signal is illustrated, which shows the consistency with the ground truth, i.e., the red stars in the figure. 
    }\label{fig:vali_D_t_plot}
\end{figure}

Utilizing a walking scenario as shown in Fig.~\ref{fig:scenario} (a) for simulation. The detailed geometry information is illustrated in Fig.~\ref{fig:scenario} (b). A human walks circularly at a constant speed and the radius of the circle is 3 meters.
To observe the obvious change in the velocity, the walking speed may not be realistic in indoor scenarios, where the radius of the circle is 3 meters and the number of animation frames is set to be $100$ with a frame rate of $24$ Hz. Hence the walking speed along the tangent of the circle is 
\begin{equation}
    \begin{aligned}
        S_\text{walk} = \arctan{\dfrac{2 \times 3 \times \pi}{100 \times \dfrac{1}{24} }}= 4.52 \quad  m/s.
        \end{aligned}
\end{equation}
In the simulation, the walking direction is not towards the AoA direction, therefore the detected velocity of the main human body is less than the $S_{walk}$. 

Utilizing the FFT on the simulated data cube of each CPI, we can obtain the range-Doppler map. Collecting the velocity spectrum with the maximum power of each CPI, the concatenated Doppler-time results are obtained in Fig.\ref{fig:vali_D_t_plot}, where the continuous changing of Doppler velocities alone with time is observed in Fig \ref{fig:vali_D_t_plot}.

The mean error of the estimated velocity by simulation and the ground truth is calculated as 
\begin{equation}
    \begin{aligned}
        E_\text{velo} =  \dfrac{ \Sigma_{i = 1}^{N} \mid  v_\text{est} -  v_\text{tru} \mid }  {N_p}  = 1.3 \quad  m/s,
        \end{aligned}
\end{equation}
where $N$ is the number of CPIs, $i$ is the index of the CPI, $v_\text{tru}$ is the ground truth, and $v_\text{est}$ is the simulated velocity corresponding to the maximum value of each range-velocity spectral. 

\section*{Acknowledgment}
This work was supported by the Luxembourg National Research Fund (FNR) through the BRIDGES project MASTERS under grant BRIDGES2020/IS/15407066.

\bibliographystyle{IEEEtran}
\bibliography{bib.bib}
\end{document}